\documentclass[aps,prd,nopacs,nofootinbib,notitlepage]{revtex4-1}

\usepackage{amsfonts}
\usepackage{amsmath}
\usepackage{xcolor}
\usepackage{bm}
\usepackage{fancybox}
\usepackage{comment}
\usepackage{multirow}
\usepackage{tipa}
\usepackage{longtable}
\definecolor{refs}{RGB}{245,156,74}
\usepackage[colorlinks=true,hyperfootnotes=true,citecolor=cyan]{hyperref}

\usepackage{enumitem}

\newcommand{\be}{\begin{equation}}
\newcommand{\ee}{\end{equation}}
\newcommand{\ba}{\begin{eqnarray}}
\newcommand{\ea}{\end{eqnarray}}
\newcommand{\bs}{\begin{subequations}}
\newcommand{\es}{\end{subequations}}

\newcommand{\bbe}{\boldsymbol{\mathrm{e}}}

\newcommand{\bbie}{\boldsymbol{\textbf{\textschwa}}}

\newcommand{\bdelta}{\boldsymbol{\delta}}

\newcommand{\blambda}{\boldsymbol{\lambda}}

\newcommand{\bTheta}{\boldsymbol{\Theta}}

\newcommand{\bbell}{\boldsymbol{\ell}}

\newcommand{\bA}{\boldsymbol{A}}

\newcommand{\bC}{\boldsymbol{C}}

\newcommand{\bE}{\boldsymbol{E}}

\newcommand{\bG}{\boldsymbol{G}}
\newcommand{\bh}{\boldsymbol{h}}
\newcommand{\bJ}{\boldsymbol{J}}
\newcommand{\bj}{\boldsymbol{j}}
\newcommand{\bL}{\boldsymbol{L}}
\newcommand{\bP}{\boldsymbol{P}}

\newcommand{\bQ}{\boldsymbol{Q}}
\newcommand{\bq}{\boldsymbol{q}}
\newcommand{\bR}{\boldsymbol{R}}
\newcommand{\bT}{\boldsymbol{T}}
\newcommand{\bt}{\boldsymbol{t}}

\newcommand{\bX}{\boldsymbol{X}}

\newcommand{\bY}{\boldsymbol{Y}}
\newcommand{\by}{\boldsymbol{y}}

\newcommand{\volume}{\boldsymbol{\left(}\ast 1\boldsymbol{\right)}}
\newcommand{\inertial}{\hat{=}}

\newcommand{\bdiff}{\boldsymbol{\mathrm{d}}}
\newcommand{\bDiff}{\boldsymbol{\mathrm{D}}}

\newcommand{\lp}{\left(}
\newcommand{\rp}{\right)}
\newcommand{\lb}{\left[}
\newcommand{\rb}{\right]}

\newcommand{\nn}{\nonumber}

\newcommand{\1}{1$^\text{st}$}
\newcommand{\2}{2$^\text{nd}$}
\newcommand{\3}{3$^\text{rd}$}

\begin{document}

\title{Integrable gravity with boundaries}

\author{Tomi S. Koivisto}
\address{Laboratory of Theoretical Physics, Institute of Physics, University of Tartu, W. Ostwaldi 1, 50411 Tartu, Estonia}
\address{National Institute of Chemical Physics and Biophysics, R\"avala pst. 10, 10143 Tallinn, Estonia}
\email{tomi.koivisto@ut.ee}

\begin{abstract}

Symmetric teleparallel gravity is shown to be integrable in the presence of boundaries, given the consistent 
implementation of constraints in the covariant phase space formalism.

\end{abstract}

\maketitle

\section{Introduction}

Teleparallelism is defined by $\bDiff^2=0$, where we may consider $\bDiff$ as the exterior covariant derivative of any General Linear group GL($n$) in any $n$ dimensions.
It has been argued that spacetime is {\it a priori} described by such a trivial algebra \cite{Koivisto:2018aip}, and that the resulting theory of gravity
improves General Relativity (GR) by introducing a principle of relativity with the properly defined (equivalence classes of) inertial frames \cite{Koivisto:2022nar}. 
Besides the freedom to use arbitrary coordinates, the dynamics of the theory are equivalent in arbitrary {\it gravitational frames}\footnote{The dynamical/{\it frame symmetry} only requires that the action $I$ in Eq.(\ref{action}) is a scalar, whereas the redundancy/{\it gauge symmetry} implies an (n,1)-form identity obtained from Eq.(\ref{variation}) and stating the covariance of the density $\bL$.}. In particular, in an inertial frame \cite{BeltranJimenez:2019bnx} the field equations read $\bDiff\bh_a \inertial\bt^M_a$, where $\bh_a$ is the excitation and $\bt^M_a$ is the material energy current \cite{Koivisto:2019ggr}. Recently, $\bh_0$ was found to emerge as the Noether-Wald potential of the theory\footnote{And further,
it was shown to be the local and covariant generalisation (obtained by minimal coupling) of a variety of energy complexes, introduced by Bergmann-Thomson, von Freud, Landau-Lifshitz, Papapetrou and Weinberg \cite{Gomes:2022vrc}.} \cite{Iyer:1994ys,BeltranJimenez:2021kpj,Heisenberg:2022nvs}. However, the relation of $\oint\bh_0$ to the Hamiltonian charge was not yet rigorously clarified. 

A related issue is the integrability of the charges, and more generally, the consistency of the action principle in the presence of boundaries. 
This is the source of major technical complications in the conventional formulations of GR \cite{Regge:1974zd}. The standard action requires, without clear physical interpretation, a somewhat ambiguous piece which is confined to the boundary hypersurface. The issue is not avoided simply by turning to a \1 order formulation, 
since the on-shell action should be stationary also when variations may not vanish at the boundary. If the density $\bL=0$ at the boundary, 
the variation $\delta \int \bL$ is unproblematical. However, the conventional method is to cancel some of the unwanted terms by inserting a surface term to the action and to eliminate the remaining unwanted terms by restricting only to transformations adapted to the particular boundary geometry \cite{PhysRevLett.28.1082,Gibbons:1976ue}. The possible new method we propose is based on the frame-dependence of the density $\bL$ in symmetric teleparallelism. We may adjust the frame according to quite arbitrary boundaries, such that the bounded action is manifestly differentiable wrt arbitrary transformations.   

In section \ref{cps} some basics of the covariant phase space formalism are introduced and extended to incorporate the frame transformations.  
In section \ref{stg} the formalism is adapted to symmetric teleparallelism, and applied to both the gauge and the frame transformations. A simple example is worked out explicitly in section \ref{cgr}, and section \ref{concl} is the brief conclusion.  

\section{Covariant phase space}
\label{cps}

Let us be given some generic boundary $\mathcal{B}$. (For concreteness, say we have a manifold $\mathcal{M}$ whose boundary $\partial\mathcal{M} = \mathcal{I}^-\cup\mathcal{B}\cup\mathcal{I}^+$ consists of a spatial part $\mathcal{B}$ that we're interested in, and some past and future boundaries $\mathcal{I}^\pm$.) 

Consider an action
\be \label{action}
I = \int_{\mathcal{M}}\bL + \int_{\partial\mathcal{M}} \bbell\,,
\ee
where $\bL$ is a $(n,0)$-form and $\bbell$ is a $(n-1,0)$-form. The variation of $\bL$ can always be written as 
\be \label{variation}
\delta \bL = \bE_A\delta\phi^A + \bdiff\bTheta\,,
\ee 
where $\bE_A = 0$ are the Eqs of motion for the fields $\phi^A(x)$ and the $(n-1,1)$-form  $\bTheta$ is the pre-symplectic potential. Stationarity of the action $\delta I = 0$ (up to the future and past boundary terms) requires
\bs
\ba 
\bE_A\delta\phi^A & = & 0\,, \\
\lp \bTheta + \delta\bbell\rp\rvert_{\mathcal{B}} & = & \bdiff \bC\,, \label{bC}
\ea  
\es
where $\bC$ is a $(n-2,1)$-form (which we can arbitrarily choose due to the so-called $\bY$-ambiguity) \cite{Harlow:2019yfa}. Then the pre-symplectic form ${\Omega}'$ is defined as an integral of the pre-symplectic $(n-1,2)$-current as 
\be \label{omega}
{\Omega}' = \int_{\mathcal{C}}\delta \lp \bTheta - \bdiff\bC\rp\,,
\ee
over a Cauchy slice $\mathcal{C}$.

A symplectic manifold is defined as the pair of an abstract space $\mathcal{P}$ and a closed and non-degenerate symplectic form $\Omega$. In a covariant Hamiltonian approach, one considers the phase space $\mathcal{P}$ of field configurations, such that the $T\mathcal{P}$ is spanned by the differentials $\delta\phi^A$. The idea is that the space of solutions (wherein one usually works in the non-covariant Dirac-Bergmann Hamiltonian formalism) should be isomorphic to the space of all valid initial data on a given $\mathcal{C}$, and can therefore be considered in a covariant fashion in terms of the $\mathcal{P}$. The space of stationary field configurations is called the prephase space $\mathcal{P}'$, and the phase space is obtained by quotienting by the gauge symmetry. It is for this reason the $\Omega'$ in (\ref{omega}) was called the presymplectic form. A non-degenerate $\Omega$ is obtained as its quotient. 

However, here we only intend to suggest some \1 steps towards the covariant phase space analysis of teleparallel theories, and will not focus on Eq.(\ref{omega}) but on the boundary condition Eq.(\ref{bC}).

\subsection{Gauge transformation}
\label{gauge}

Next, consider a Diff($n$) generated by $\xi$. Define, for an arbitrary $(i,j)$-form $\bX[\delta \phi,\dots] $, the 2 contractions,  
the usual $\xi\lrcorner\bX$ which results in an $(i-1,j)$-form, and $\xi\cdot\bX[\delta \phi,\dots] = \bX[\pounds_\xi \phi, \dots]$ which results in an $(i,j-1)$-form. 
The latter contraction can be understood in terms of the vector $V_\xi$ defined in the configuration space such that
\be
V_\xi = \int \bdiff^n x \pounds_\xi \phi^A(x)\frac{\delta}{\delta \phi^A}\,, \quad \Rightarrow \quad
\bdiff\lp \xi\cdot\bX\rp  =  \xi\cdot\bdiff\bX\,, \quad \delta\lp \xi\lrcorner\bX\rp  =  \xi\lrcorner\delta\bX\,, \quad \delta_\xi\delta\phi^A(x)  =  \delta\lp \delta_\xi \phi(x)\rp\,. 
\ee
In this notation, the familiar Noether current $(n-1,0)$-form can be expressed as
\be
\bJ_\xi = \xi\cdot \bTheta - \xi\lrcorner\bL\,. 
\ee
A central result in the covariant phase space formalism is the formula for the Hamiltonian \cite{Harlow:2019yfa} (up to an irrelevant constant) that generates the family of Diffs,
\be \label{hamiltonian}
H_\xi =  \int_{\partial\mathcal{C}}\lp \bj_\xi + \xi\lrcorner\bbell - \xi\cdot\bC\rp\,,
\ee 
where the $(n-2,0)$-form $\bj_\xi$ is the Noether-Wald potential which can always be locally found such that $\bJ_\xi = \bdiff\bj_\xi$ \cite{Iyer:1994ys}. Using (\ref{bC}), one can check that the Hamiltonian is independent of the choice of $\mathcal{C}$. 

The Diff($n$) gauge transformation could be called a total covariance\footnote{General coordinate invariance has been regarded an ``improper'' symmetry (Noether), and coordinatisations in general ``a formal scaffolding'' (Weyl) to be discarded at a later stage. We will arrive at the result that the Diff($n$) is a {\it trivial symmetry} according to Freidel {\it et al}'s  \cite{Freidel:2020xyx} definition $H_\xi = 0$ {\it i.e.} not even a surface Hamiltonian survives on shell. (In this particular case, the Diff($n$) may be technically regarded a ``fake symmetry'', achieved
via St\"uckelbergisation/Kretchmannisation \cite{BeltranJimenez:2022azb}).}, since the passive operation in $\mathcal{M}$ is made tautologically active, and we may write e.g.
$\delta_\xi \phi^A(x) = \pounds_\xi \phi^A(x) = \{ \bDiff, \xi\lrcorner\} \phi^A(x) \overset{}{=} V_\xi\lrcorner \delta \phi^A(x) = \pounds_{V_\xi}\phi^A(x) = \{ \delta, \xi\cdot\} \phi^A(x)$. 


\subsection{Frame transformation}
\label{frame}

Previous works on Hamiltonian analysis of $[\nabla,\nabla]=0$ gravity have proven that the $\mathcal{P}$ for actions (\ref{action}) can be well defined \cite{DAmbrosio:2020nqu,Hu:2022anq}. (Since by fixing the GL($n$)-invariance of the density $\bL_Q$ we shall introduce at Eq.(\ref{LQ}) to the coordinate frame $\bbe^a=\bdelta^a$ in Eq.(\ref{LQcgr}) it becomes the density of the Coincident GR \cite{BeltranJimenez:2017tkd} and by fixing further the gravitational frame imposing the coincident gauge this density $\bL_Q$ becomes the same $\bL_{ADM}$ used in the standard ADM Hamiltonian treatment, it is clear that the Cauchy problem can be well-posed, and that we recover the $\mathcal{P}$ of GR.)

Here our approach is completely different, since the aim is to take into account the frame-dependence of the symplectic structure, the Hamiltonian and other charges {\it etc}.  
In the case that an $I$ involves background fields, some of the nice identities at the end of \ref{gauge} are violated. We have demanded the covariance of each $\phi^A$ under a gauge-Diff($n$), but this will not be the case under a frame-$\tilde{\text{Diff}}$($n$). The latter can indeed be interpreted as transformations which leave some of the fields frozen into the role of background fields. It is convenient to realise a frame transformation as a reconfiguration of the affine structure of $\mathcal{M}$.

The suggested construction of $\mathcal{P}$ proceeds as follows. 
\begin{itemize}
\item[(1)] Assume an \underline{invariant functional $I$ of torsor connection}\footnote{This means that we can freely translate the connection. The translation invariance can be straightforwardly generalised to full connection-independence, extending the symmetry to GL($n$)$\times\tilde{\text{GL}}$($n$) \cite{BeltranJimenez:2019odq}.}. The symmetry is then GL($n$)$\times\tilde{\text{Diff}}$($n$).   
\item[(2)] Imposing $\delta I =0$, and in particular symmetric teleparallelism\footnote{$\bDiff^2=0$ is the dynamical consequence of $m_P$ being the mass of the connection \cite{Koivisto:2019ggr}, as the density $\bL_Q$ clearly suggests. However, restricting here to sub-Planckian scales, we need not consider an explicit kinetic term for the connection, but will set $\bDiff^2=0$ using multipliers.}, the symmetry is reduced to Diff($n$)$\times\tilde{\text{Diff}}$($n$). We are then looking at some meta-phase space $\mathcal{P}''$, accommodating physically distinct classes of theories and their gauge-degenerate field configurations. The same configuration $\{\phi^A(x)\}$ can represent many different theories, since the mappings from a configuration to the observables are frame-dependent.
\item[(3)] Imposing $\bL\vert_{\mathcal{B}}=0$, the frame is fixed such that the action principle is well defined in the presence of the given boundary $\mathcal{B}$. If these boundary conditions determine the frame completely, the remaining symmetry is Diff($n$). This is the prephase space $\mathcal{P}'$.
\item[(4)] The non-degenerate phase space $\mathcal{P}=\mathcal{P}'/$Diff($n$) is well-defined, though in general inequivalent to the  $\mathcal{P}$ of GR.
\end{itemize}
The application of this schema will be illustrated, after setting up the general formalism in section \ref{stg}, with a simple example in section \ref{cgr}.
 
\section{Symmetric teleparallelism}
\label{stg}

The fields $\phi^A$ in listed in table \ref{fields} below are conventional in metric-affine gravity. 
\begin{center}
\begin{table}[h]
\begin{tabular}{| l c | c | l    l  | |   l | }
\hline
 the metric  & $(0,0)$-form & $g_{ab}$  & $\Rightarrow$ nonmetricity & $\bQ_{ab}  = \bDiff g_{ab}$ & $= \bdiff g_{ab} + 2\bA_{ab}$ \\  \hline
 the $n$-bein  & $(1,0)$-form & $\bbe^a$  & $\Rightarrow$ torsion & $\bT^a = \bDiff \bbe^a$ & $= \bdiff \bbe^a - \bA_b{}^a\wedge\bbe^b$ \\  \hline
 the connection  & $(1,0)$-form & $\bA_a{}^b$  & $\Rightarrow$ curvature & $\bR_a{}^b = \lp \bDiff \bA\rp_a{}^b$ & $= \bdiff \bA_a{}^b + \bA_a{}^c\wedge\bA_c{}^b$ \\  \hline
\end{tabular}
\caption{The generic set of gravitational fields $\phi^A$. \label{fields}}
\end{table}
\end{center} 
If we set symmetric $\bT^a=0$ teleparallelism $\bR_a{}^b=0$ with $(n-2,0)$-form multipliers, those formally count as yet additional fields. The action would be
\be \label{Qaction}
I = \int \bL = \int \lp \bL_Q + \blambda^a{}_b\wedge\bR_a{}^b + \blambda_a\wedge\bT^a\rp \,, 
\ee
wherein the $\bL_Q$ is responsible for the dynamics of non-metricity. It is useful to also define
\bs
\label{momenta}
\ba
\text{the non-metricity conjugate :} \quad \bq_{ab}  & = & \frac{\partial\bL_Q}{\partial\bQ^{ab}}\,, \label{conjugate} \\
\text{the metric energy current :} \quad \bG_{ab}  & = &  -\frac{\partial\bL}{\partial g^{ab}} -\bQ_a{}^c\wedge\bq_{cb}\,, \label{mconjugate}\\
\text{the $n$-bein energy current :} \quad \bt_a  & = & - \frac{\partial\bL}{\partial \bbe^{a}}\,. \label{econjugate}
\ea
\es
Thus, we have the density
\be  \label{LQ}
\bL_Q = \frac{1}{2}\bQ^{ab}\wedge\bq_{ab}\,,
\ee 
with some generic $\bq_{ab}$. 
Variations then yield us
\bs
\label{variations}
\be
\bE_A\delta\phi^A  =  \delta g^{ab}\lp g_{ac}\bDiff\bq^c{}_b - \bG_{ab}\rp  + \delta\bbe^a\wedge\lp \bDiff\blambda_a-\bt_a\rp 
+\delta\bA_a{}^b\wedge\lp 2\bq^a{}_b - \bbe^a\wedge\blambda_b + \bDiff\blambda^a{}_b\rp + \delta\blambda^a{}_b\wedge\bR_a{}^b + \delta\blambda_a\wedge\bT^a\,, 
\ee
and the symplectic current
\be
\bTheta = -\delta g^{ab}\bq_{ab} + \delta\bbe^a\wedge\blambda_a + \delta \bA_a{}^b\wedge\blambda^a{}_b\,.
\ee
\es
The Eqs of motion $\bE_A\delta\phi^A=0$ imply the 3 Eqs,
\bs
\label{equations}
\ba
2\bG^a{}_b & = & -\bbe^a\wedge\bt_b \quad \Rightarrow \quad \bt_a = -2\bbie_b\lrcorner\bG^b{}_a\,, \label{structure1} \\ 
\bDiff\lp\bbie_a\lrcorner\bDiff\bq^a{}_b\rp & = & 0  \quad \Rightarrow \quad  \mathring{\bDiff}\ast\bt_a^M = 0\,, \label{structure2} \\
2\bDiff\bq^a{}_b & = & -\bbe^a\wedge\bDiff\blambda_b  \quad \Rightarrow \quad \blambda_a = \bh_a + \bDiff \by_a  \label{structure3}\,.
\ea
\es
The \1 Eq. shows that the metric and the $n$-bein inertiality criterions, $\bG_{ab} \inertial0$ and $\bt_a \inertial0$ respectively, are equivalent. The \2 Eq. is the Bianchi identity of the frame invariance\footnote{This would generically be broken by modifications of GR introducing new degrees of freedom. Such could render Eq.(\ref{structure2}) only an on-shell
identity, potentially spoiling the isomorphism of the space of solutions and the space $\mathcal{P}$. The question whether there exists a ``properly parallelised'' frame \cite{Koivisto:2018loq} (wherein the rank of $\Omega$ would be a constant), is outside our scope here since the starting point (1) as stated in section \ref{frame} excludes the modifications of GR.}.
The \3 Eq. can be used to determine the excitation $\bh_a$, and we have parameterised the arbitrariness of the solution with a $(n-3,0)$-form $\by_a$. 

In the end the full dynamics, taking into account the possible material current $\bt^M_a$, are described by the gauge-covariant and frame-invariant field Eq. $\bDiff\bh_a = \bt^M_a + \bt_a$, .

\subsection{Gauge transformation}

The presymplectic potential of a Diff $\xi$ is
\ba
\xi\cdot\bTheta  & = &  \lp\xi\lrcorner\bQ^{ab}\rp\bq_{ab} + \lp \xi\lrcorner\bT^a + \bDiff\xi^a \rp \wedge \blambda_a +\xi\lrcorner\bR_a{}^b\wedge\blambda^a{}_b \nn \\
 & = &  \bdiff\bj_\xi + \xi\lrcorner\bL -\xi^a\bE_a + \bT^a\wedge\xi\lrcorner\blambda_a  + \bR_a{}^b\wedge\xi\lrcorner\blambda^a{}_b\,,
\ea 
where
\bs
\ba
\xi\lrcorner\bL & = & -\xi^a\bt_a + \lp\xi\lrcorner\bQ^{ab}\rp\bq_{ab} + \xi\lrcorner\lp \bT^a\wedge\blambda_a + \bR_a{}^b\wedge\blambda^a{}_b\rp\,, \\
\bj_\xi & = & \xi^a\bh_a\,. \label{disc}
\ea
\es 
In the above we discarded the $\by$-ambiguity from (\ref{structure3}) since the integral of an exact form over a closed surface $=0$ and therefore the $\by$ does not contribute to the observable 
charges\footnote{This does not quite account for the difference of the Noether-Wald potentials obtained in the Palatini \cite{Heisenberg:2022nvs} and in the GL($n$) \cite{BeltranJimenez:2021kpj} formulations since it is not exact but instead $\Delta \bj_\xi = m_P^2\ast\bdiff{}^\flat{}\xi$ \cite{Gomes:2022vrc}.}. We see that the stationarity condition (\ref{bC}) holds (neglecting an irrelevant $\by$-type and $\bY$-type ambiguity in $\bC$) if
\bs
\ba
\xi\lrcorner\bL\rvert_{\mathcal{B}} & = & 0\,, \label{bc1} \\
\xi\cdot\bC\rvert_{\mathcal{B}} & = &  \xi^a\bh_a\,. \label{bc2}
\ea
\es
We can satisfy (\ref{bc1}) with an arbitrary $\xi$ by adjusting the frame of a configuration such that the density $\bL$ vanishes at the given $\mathcal{B}$. This condition boils down to setting 1 scalar function $L(x \in \mathcal{B})=0$ to vanish, and thus we can safely assume a solution to exist. At least locally, the density-free frame can also be an inertial frame
(point-wise, one can simply find a freely-falling coordinate system in the coincident gauge). 
The density-free boundary condition agrees with the intuition extrapolated from the case which is best understood in GR, a flat $\mathcal{B}$ at spatial infinity. It is clear that 
asymptotically far away from all the sources, the density $\bL$ may grow instead of decay only wrt some kind of non-inertial reference frame, and this has indeed
been considered in the context of teleparallel gravity as a criterion for regularised energy expressions and actions \cite{Lucas:2009nq, Krssak:2015lba}. The no-density consistency condition (\ref{bc1}) applies to a generic boundary, and thus provides the local and covariant generalisation of the physically acceptable boundary conditions for a flat $\mathcal{B}$
at infinity. Finally, we note that the identification (\ref{bc2}) vanishes the Hamiltonian (\ref{hamiltonian}), $\mathcal{H}_\xi = 0$. 
\subsection{Frame transformation}

A transformation which leaves the density $\bL$ invariant only up to an exact form is sometimes called a pseudosymmetry.
A remarkable property of Coincident GR is the pseudosymmetry wrt independent translations of the connection \cite{BeltranJimenez:2017tkd}, 
\be
\Delta \bL = \bdiff\lp2\Delta\bA_a{}^b\wedge\bq^a{}_b\rp\,.
\ee
On the other hand, by adapting (\ref{variations}) we obtain, in the variation (\ref{variation}) 
\bs
\be \label{variation2}
\xi\lrcorner\Delta \bL = \Delta\bA_a{}^b\wedge\lp 2\bq^a{}_b-\bbe^a\wedge\blambda_b + \bDiff\blambda^a{}_b\rp + \bdiff\lp \Delta\bA_a{}^b\wedge\blambda^a{}_b\rp\,,
\ee
and the presymplectic potential
\be
\bTheta = \bDiff\lp\bbie_a\lrcorner\bDiff\xi^b\rp\wedge\blambda^a{}_b = \bbie_a\lrcorner\bDiff\xi^b\wedge\lp  \bbe^a\wedge\bh_b - 2\bq^a{}_b\rp\,,
\ee
\es
where the \2 form follows by discarding a term that does not contribute to the variation (\ref{variation2}). We now find the Noether current
\bs
\be
\bJ_\xi = \bDiff\xi^a\wedge\bh_a {\inertial}   \bdiff\lp \xi^a\bh_a\rp\,.
\ee
The hatted equality assumed an inertial frame, $\bt_a \inertial 0$. Thus, the Noether charges of the frame (pseudo)symmetry and the gauge symmetry
are the same in an inertial frame but not otherwise. Furthermore, the quasi-local Noether charge is also the Hamiltonian generator of the frame transformation,  
\be
H_\xi = -\int_{\mathcal{C}}\bDiff\xi^a\wedge\bt_a + \int_{\partial \mathcal{C}} \xi^a\bh_a \inertial\int_{\partial \mathcal{C}}\xi^a\bh_a\,. 
\ee 
\es
Transition into a non-inertial frame $\bt^a \neq 0$ can generate a bulk Hamiltonian.  

\section{Coincident General Relativity}
\label{cgr}

We shall make explicit the relation between the symmetric $\bDiff^2=0$ version of GR originally suggested by Nester \& Yo \cite{Nester:1998mp,Adak:2005cd} and the $[\nabla,\nabla]=0$ version of GR introduced by Beltr{\'a}n {\it et al} \cite{BeltranJimenez:2017tkd,BeltranJimenez:2018vdo}. To make contact with the Palatini (tensor) formulation, we define the contravariant vectors 
\bs
\ba
\bQ & = & g^{ab}\bQ_{ab}\,, 
\\  
\tilde{\bQ} & = & \lp\bbie_a\lrcorner\bQ^a{}_b\rp\bbe^b\,.  
\ea
The special case of a density (\ref{LQ}) we consider in this section is   
\es
\ba
m_P^{-2}\bL_Q &=&  -\frac{1}{8}\bQ_{ab}\wedge\ast\bQ^{ab}  + \frac{1}{4}\bQ_{ac}\wedge\bbe^a\wedge\ast\lp \bQ^{bc}\wedge\bbe_b\rp - \frac{1}{8} \bQ\wedge\ast\bQ + \frac{1}{4}\bQ\wedge\ast\tilde{\bQ}  \nn \\
& = &  \frac{1}{2}\lp -\frac{1}{4}Q_{abc}Q^{abc} + \frac{1}{2}Q_{abc}Q^{bac} + \frac{1}{4}Q_a Q^a - \frac{1}{2}Q_a\tilde{Q}^a\rp\lp \ast 1\rp\,.  \label{LQcgr}
\ea
\bs
The non-metricity conjugate (n-1,0)-form (\ref{conjugate}) derived for (\ref{LQcgr}) is
\ba
m_P^{-2}\bq_{ab} & = &  -\frac{1}{4}\ast\bQ_{ab} + \frac{1}{2}\bbe_{(a}\wedge\ast\lp\bQ_{b)c}\wedge\bbe^{c}\rp - \frac{1}{4}g_{ab}\lp \ast\bQ - \ast\tilde{\bQ}\rp
+ \frac{1}{4}\bQ_{(a}\lp\ast\bbe_{b)}\rp \nn \\
& = & -\frac{1}{4}\lb - Q^{c}{}_{ab} + 2Q_{(ab)}{}^c + g_{ab}\lp Q^c - \tilde{Q}^c \rp - Q_{(a}\delta^c_{b)} \rb \lp \ast \bbe_c\rp\,. 
 = -m_P^{-2}\ast \bP_{ab}\,. 
\ea
In the last step we borrowed a notation for the 1-form $\bP_{ab} = \ast\bq_{ab}$ from the tensor formalism. 
Now we solve the excitation from (\ref{structure3}), 
\be \label{excitation}
\bh_a =  m_P^2 \ast \lb \bQ_{ab}\wedge\bbe^b + \bbe_a\wedge\lp \bQ-\tilde{\bQ}\rp\rb\,. 
\ee
The $n$-bein energy current (\ref{econjugate}) is
\ba
\bt_a & = & -\bbie_a\lrcorner\bL_Q + \frac{m_p^4}{4}\lb \bbie_a\lrcorner\bQ_{bc}\lp 2\bbe^b\wedge\ast\lp\bQ^c{}_d\wedge\bbe^d\rp +\bbie^{(b}\lrcorner\bQ\ast\bbe^{c)}-\ast\bQ^{bc}\rp
+ \bbie_a\lrcorner\bQ\lp\bbie_b\lrcorner\tilde{\bQ}\ast\bbe^b - \ast\bQ\rp\rb \nn \\
& = & \lp - \frac{1}{2}\delta^b_aQ_{cde}P^{cde} + Q_a{}^{cd}P^b{}_{cd}\rp\lp\ast \bbe_b\rp\,.
\ea
We compute also the metric energy current (\ref{mconjugate}), and as a cross-check verify the identity (\ref{structure1}),
\be
2\bG^a{}_b  = \delta^a_b\bL_Q - \bbie_b\lrcorner\bQ^{cd}\bP_{cd}\wedge\ast\bbe^a 
 =  \lp\frac{1}{2}\delta^a_b  Q_{cde}P^{cde} - Q_b{}^{cd}P^a{}_{cd}\rp\volume  =   -\bbe^a\wedge\bt_b\,. 
\ee 
\es
The results of the analysis are summarised in table \ref{taabeli}. 
\begin{center}
\begin{table}
\begin{tabular}{ |c|c|c| c |} 
 \hline
 form & GR: gauge & CGR: gauge & CGR: frame \\ 
 \hline
 density $\bL$ & Einstein-Hilbert & \multicolumn{2}{c|}{$\bL_Q=\bQ^{ab}\wedge\bq_{ab}/2$\,, c.f. Eq.(\ref{LQcgr})}  \\ 
 surface density $\boldsymbol{\ell}$ & Gibbons-Hawking-York & \multicolumn{2}{c|}{-}  \\ 
 \hline
 Noether charge $\oint\bj$ & Komar & $\oint\bh$\,, c.f. Eq.(\ref{excitation}) & $\hat{=}\oint\bh$\\
 Hamiltonian $H$ & Brown-York & $0$ & $\hat{=}\int_{\partial\mathcal{C}}\bh$  \\
 \hline
 conditions $@\mathcal{B}$    & $\xi$ is tangential Killing & $\bL=0$ & n/a \\
 \hline
\end{tabular}
\caption{Summary of our conclusions and the well-known results in GR.\label{taabeli}} 
\end{table}
\end{center}

\subsection{Example}
\label{example}

It can be useful to illustrate the role of the 2 types of transformations with a simple example. Set $n=4$ and take the cosmological solution, 
\be \label{frw}
\bdiff s^2 = -n^2(t)\bdiff t^2 + a^2(t)\delta_{ij}\bdiff x^i \bdiff x^j\,.
\ee
One may want to consider this in an $\mathcal{M}$ bounded by the cosmological horizon, or maybe to have an action bounded by a given 
event's past light cone. Such could be consistently described by a theory wherein the density $\bL$ vanishes at the boundary. 
Given the line element (\ref{frw}), the choice of frame simply corresponds to the choice of connection. 
Hohmann has constructed the most general homogeneous and isotropic symmetric teleparallel geometry  \cite{Hohmann:2021ast}, and we adopt his \1 solution\footnote{Hohmann reported 3 branches of solutions for the connection \cite{Hohmann:2021ast}, but our conclusions would be similar in the 2 other branches. The most general case of spherically symmetric geometry has also been nicely explored \cite{DAmbrosio:2021zpm,Bahamonde:2022zgj}.} characterised by $1$ free function, $K(t)$. The $\bL_Q$ depends upon this function as
\be
\bL_Q = \frac{3m_P^2}{2}\lp 2H^2 + 3HK + \dot{K}\rp\volume\,,  \quad \text{where} \quad H= \frac{\dot{a}}{a}\,, \quad \frac{d f}{dt} = n\dot{f}\,.
\ee
Given the dynamics encoded in $a(t)$, from $\bL_Q=0$ we obtain an inhomogeneous, \1 order ordinary differential Eq. to determine the function $K(t)$.
To find an explicit solution, let us add a matter source, in the simplest case a $\Lambda$-term,
\be
\bL = \bL_Q + \bL_\Lambda = \frac{m_P^2}{2}\lp 6H^2 + 9HK + 3\dot{K} + 2\Lambda\rp a^3n\bdiff^4 x \overset{n=1}{=} \frac{m_P^2}{2}\lp 4\Lambda + 3\sqrt{3\Lambda}K 
+ 3\dot{K}\rp  e^{\sqrt{3\Lambda}t}a^3_0 \bdiff^4 x
\ee 
Both the choice of the cosmological connection and the choice of the lapse is irrelevant to the dynamics, even though the former will recalibrate the energy units 
and the latter will change the interpretation of the coordinate time $t$. More generally, time-space intervals are varied in a gauge-Diff, whilst the differencies in the gauge-invariant energy-momentum charges $\oint\bh_a$ are varied in a frame-$\tilde{\text{Diff}}$. 

Let us now walk through the steps we recall from section \ref{frame}. (1) We have an invariant $\bL$ such that it changes by an exact form when translating the connection. (2) The meta-phase space $\mathcal{P}''$ corresponds to all the available solutions, including now 2 arbitrary functions $n(t)$ and $K(t)$. (3) By imposing the density-free boundary condition, we are given from the space $\mathcal{P}''$ the slice $\mathcal{P}'$ wherein $K$ is fixed such that on the n(t)=1 hyperslice of $\mathcal{P}'$ it is the constant $K=-4/3\sqrt{\Lambda/3}$ (a 1-parameter family of solutions is found iff for some $\mathcal{C} \supset \mathcal{B}$). (4) In the phase space $\mathcal{P}=\mathcal{P}''/\text{Diff}$ also the degeneracy due to $n(t)$ is eliminated, since we {\it mod} out the time-reparameterisation gauge invariance.

\section{Conclusion}
\label{concl}

Symmetric teleparallel gravity features the so-called frame pseudosymmetry. Recovering the standard ADM formulation of GR is one way to the fix the frame, but in a manifold
with a boundary $\mathcal{B}$, the well-posedness of the action principle can provide the more appropriate criterion.    
By the consistent choice of frame one may incorporate arbitrary gauge transformations in arbitrary geometry: 
\begin{itemize}
\item We see from (\ref{bc1}) that the $\xi_\perp$' s normal to $\mathcal{B}$ are automatically integrable, avoiding artificial restrictions to diffeomorph the total $\mathcal{M}$ with boundaries.     
\item The $\xi_\parallel$'s tangential to $\mathcal{B}$ require the no-density boundary condition. We emphasise the viewpoint that (\ref{bc1}) is imposed in $\mathcal{P}''$ {\it i.e.} the boundary condition is a restriction upon the resulting covariant phase space $\mathcal{P}$. 
\end{itemize}
An example in section \ref{example} demonstrated that the no-density boundary condition determines the gravitational frame at cosmological scales. It remains to be 
explored whether this could shed light on the initial conditions required for a viable inflation. At very high energies we can no longer justify the approximation $\bDiff^2=0$ and it seems possible that the frame is settled in a dynamical fashion.  
 
\begin{acknowledgments}
This work was supported by the Estonian Research Council grants PRG356 ``Gauge Gravity'' and MOBTT86, and by the European Regional Development Fund CoE program TK133 ``The Dark Side of the Universe''.
\end{acknowledgments}

\bibliography{QCPSrefs}

\end{document}